# Hyperspectral Imaging Spectroscopy of a Mars Analog Environment at the North Pole Dome, Pilbara Craton, Western Australia

ADRIAN J. BROWN[1], MALCOLM R. WALTER[1], AND THOMAS CUDAHY[1,2]

[1] Australian Centre for Astrobiology, Macquarie University, NSW 2109, Australia, corresponding email: abrown@els.mq.edu.au; web site: http://aca.mq.edu.au/abrown.htm

[2] CSIRO Division of Exploration and Mining, ARRC Centre, 26 Dick Perry Ave, Technology Park, WA , 6102 Australia

## ABSTRACT

A visible and near infrared (VNIR) to shortwave infrared (SWIR) hyperspectral dataset of the Early Archaean North Pole Dome (NPD), Pilbara Craton, Western Australia, has been analysed for indications of hydrothermal alteration. Occurrence maps of hydrothermal alteration minerals were produced. It was found that using a spatial resolution on the ground of approximately 5m and spectral coverage from 0.4 to 2.5 microns was sufficient to delineate several hydrothermal alteration zones and associated veins, including phyllic, serpentinitic and chloritic alteration. These results suggest this level of spectral and spatial resolution would be ideal for localising shallow epithermal activity, should such activity have existed, on the surface of Mars.

## INTRODUCTION

In October 2002 an airborne visible and near infrared to shortwave infrared (VNIR-SWIR or 0.4-2.5 micron) hyperspectral reflectance dataset was collected over the North Pole Dome (NPD) region of the Early Archaean East Pilbara Granite Greenstone Terrane (EPGGT).

The dataset covers over 600km$^2$ of the NPD (Figure 1), including some of the best exposed outcrops of the 3.5 Ga Warrawoona Group (Van Kranendonk 2000). This region has been the subject of extensive palaeobiogical study and has been suggested as a habitat for early life on Earth (Groves*, et al.* 1981; Schopf 1993). Hydrothermal veins at the NPD have been proposed as both a habitat of life (Ueno*, et al.* 2004; Van Kranendonk and Pirajno 2004) and a preservation mechanism for biosignatures (Dunlop and Buick 1981). The lateral extent and spatial associations of hydrothermal alteration at the NPD is thus a key factor in understanding the Early Archaean biosphere.

*Hyperspectral Missions to Mars*

The European Space Agency (ESA) Mars Express orbiter has carried the *Observatoire pour la Minéralogie, l'Eau, la Glace et l'Activité* (OMEGA) instrument (Bonello*, et al.* 2004) to Mars and has already returned a large amount of data covering the Martian surface (Bibring*, et al.* 2004). The *Compact Reconnaissance Infrared Spectrometer for Mars* (CRISM) instrument is scheduled to launch on the Mars Reconnaissance Orbiter in 2005 and return its first data to Earth in 2006. In order to interpret the spectra obtained by these instruments, Earth analog investigations are useful, as they offer the opportunity to ground truth mineral identifications and use this ground truth to refine spectral-based mapping algorithms. This project has been designed to investigate the ability of a hyperspectral survey to detect and map a hydrothermal system in a Mars analog environment.

## GEOLOGICAL SETTING

The NPD is a structural dome of bedded, dominantly mafic, weakly metamorphosed volcanic rocks (greenstones) and interbedded ultramafic, felsic and cherty horizons of the 3.51-3.43 Ga Warrawoona Group that dip 30 to 70 degrees away from the central North Pole Monzogranite, a syn-volcanic laccolith emplaced in the core of the Dome at 3.46 Ga (Thorpe, *et al.* 1992; Van Kranendonk 2000). Minor occurrences of felsic volcanic rocks are interbedded with the greenstones, and these are capped by cherts associated with hiatuses in volcanism (Van Kranendonk and Hickman 2000). Stratigraphy of the NPD is given in Table 1.

Many authors have attempted to determine the depositional setting of the Warrawoona Group. Early researchers suggested a shallow marine environment (Buick and Barnes 1984), and recent work has suggested a mid-ocean ridge (Kitajima, *et al.* 2001) or oceanic plateau (Van Kranendonk and Pirajno 2004) setting. Stromatolites and putative microfossils have been documented at three different horizons within the NPD (Dunlop, *et al.* 1978; Lowe 1980; Walter, *et al.* 1980; Awramik, *et al.* 1983; Hofmann, *et al.* 1999; Ueno, *et al.* 2004).

Hydrothermal activity at the North Pole Dome has predominantly been of a low temperature, low pressure ('epithermal') type (Barley 1984). Carbonation is broadly restricted to the mafic and ultramafic lavas of the NPD, whereas pervasive silicification events are most often associated with felsic volcaniclastic horizons and hydrothermal conduits such as those in the Dresser Formation and Strelley Pool Chert (Dunlop and Buick 1981; Lowe 1983).

This study examines a portion of the NPD dataset covering the upper Dresser Formation and parts of the Apex Basalt (Figure 1). This region was chosen due to the pervasive hydrothermal alteration present at the top of the Dresser Formation.

### *North Pole Dome as a Mars Analog*

Numerous space and planetary science reviews have highlighted the importance of continuing analysis of Earth-based analogs in order to advance the exploration of Mars and the Solar System (McCord 1988; NASA 1995; Space Studies Board 2003; Farr 2004). To this end, many studies of analog regions in volcanic, arid, biologically extreme environments have been initiated.

The NPD region is a compelling Mars analog due to features such as ancient (3.5 billion year old), well-preserved mafic-ultramafic volcanic successions, sparse vegetation, and minimal tectonic metamorphism. In addition, the presence of Earth's earliest biosignatures makes the NPD an ideal proving ground for strategies to find evidence of past life on Mars (Brown, *et al.* 2004).

## AIRBORNE IMAGING SPECTROSCOPY

Imaging spectroscopy has been used for more than 20 years to assist economic geologists in remote exploration (Goetz, *et al.* 1983). Numerous recent studies have investigated the potential for spectral imaging to map mineral assemblages remotely (Kokaly, *et al.* 1998; Stamoulis, *et al.* 2001; Bierwirth, *et al.* 2002; Kirkland, *et al.* 2002; Thomas and Walter 2002; Cudahy 2004; Hellman and Ramsey 2004; Rowan, *et al.* 2004). A large collection of papers from the annual Airborne Visual and Infrared Imaging Spectrometer (AVIRIS) workshops is archived at the Jet Propulsion Laboratory (JPL) AVIRIS website (http://aviris.jpl.nasa.gov/).

*Physical Effects*

When photons impinge upon a surface, they may be transmitted, reflected, or absorbed (Wendlandt and Hecht 1966). Reflectance spectroscopy is the measurement (over a range of wavelengths) of the flux of photons reflected from a surface in order to obtain diagnostic information about the material under study. It is a particularly useful technique for spacecraft studying large areas of planetary bodies illuminated by the sun, since there is no requirement to carry onboard a powerful illumination source (McCord 1988).

When panchromatic light interacts with a crystal, it is absorbed in anomalous amounts at particular wavelengths. The regions of anomalous absorption depend on the crystal structure and can be diagnostic for a particular mineral. However, in addition to absorption due to crystal effects, reflectance spectra can be modified by varying grain size, temperature, pressure and mixing (Singer and Roush 1985; Mustard and Hays 1997; Kirkland*, et al.* 2003). These effects can complicate spectral identification algorithms.

Because photons do not penetrate far into solid surfaces, depending on the overall absorption strength of the target, the technique of reflectance spectroscopy is only able to detect minerals within roughly 10 microns of the exposed surface of a rock (Vincent and Hunt 1968). Thick dust, desert varnish, weathering rinds, solids, sand and alluvial coverings can inhibit (or prohibit) the detection of target minerals. The influence of weathering in outback Australia is an obstacle to determination of igneous precursors, as it is in any other geochemical project.

The effects of 'mixing' due to multiple crystal configurations within the instrument's Ground Instantaneous Field Of View (GIFOV, ie. 1 pixel on the ground) has traditionally been analysed in two ways. If the mixing components under consideration are large and well separated (such as two different lava flows), the mixing can be approximated as a linear additive process ('linear mixing'), where the contribution of each material to the spectrum is proportional to its relative abundance and Beer's law applies (Clark and Roush 1984). However, if the components under consideration are intimately mixed, such as two components of a bimodal lava flow, or different sized crystals of the same chemical structure (Bishop*, et al.* 2003), the mixing can be highly non linear and the components may not be separable in a quantitative sense. The mixing of materials can make some minor components extremely difficult or impossible to recognise, especially minor carbonate phases (Pontual 1997; Kirkland*, et al.* 2001).

Minerals of interest in this study are listed in Table 2, along with their chemical formulae (Deer*, et al.* 1992) and typical hydrothermal alteration classes (Thompson and Thompson 1996). These minerals have been chosen as targets since they are detectable in the SWIR (2-2.5 micron) region due to the hydroxyl ($OH^-$) in their mineral structure. These minerals are also typically found in hydrothermal alteration zones, often in concentric patterns surrounding hot spots (Meyer and Hemley 1967; Lowell and Guilbert 1970).

Detection of minerals using spectroscopy relies on matching the spectra collected to a characteristic feature present in a library spectrum. Several spectral libraries are publicly available and this study primarily used the United States Geological Survey (USGS) Spectral Library version 4. A new version of the library has recently been made available (Clark*, et al.* 2003).

Given the amount of information present in the spectra of a mineral, it stands to reason that more information will be available in high resolution spectra (Clark, *et al.* 1990). As instrument technology has advanced, the progression from so called multispectral instruments such as the seven channel LANDSAT series, to the modern day AVIRIS instrument with 256 channels, has meant that 'laboratory quality' spectra are now available to the airborne or orbital imaging spectroscopist. The term 'hyperspectral' is typically used to define an instrument that has in excess of 60 channels. The HyMap instrument, with 126 channels, qualifies as a hyperspectral instrument

*The HyMap Instrument*

Integrated Spectronics (ISPL – http://www.intspec.com) has developed a series of hyperspectral VNIR-SWIR imaging systems which are commercially flown by HyVista Corporation (http://www.hyvista.com). This instrument series, called 'HyMap', has been under continuous development since 1998 (Cocks, *et al.* 1998).

While there are many configuration options for the HyMap instrument, the instrument was outfitted for this project with 126 spectral channels covering the VNIR-SWIR wavelengths between 0.45 and 2.5 microns. The coverage of this part of the spectrum is almost contiguous; however coverage is deliberately sparse in areas of high atmospheric absorption near 1.4 and 1.9 microns. Table 3 details the operating characteristics of the HyMap instrument for this project.

A high signal to noise ratio (SNR) is vital to provide high quality spectra that can be confidently interpreted. For imaging spectrometers, the SNR is strongly influenced by the number of photons arriving at the instrument over any part of the spectrum (Schott 1996). Diminishing numbers of photons are available at wavelengths higher than 0.5 microns due to the nature of the solar response function. Parts of the spectrum affected by atmospheric absorption will also allow fewer photons to reach the instrument, thus decreasing the SNR. A plot of the measured SNR for the HyMap instrument is given in Figure 2.

At the HyMap operating height of around 2400m above ground level (AGL), some turbulence can be experienced. HyMap is mounted on a 3-axis gyro stabilized platform with an inertial navigation unit and GPS synchronized for recording actual location in time and space as the instrument is operating. This positional data is provided by HyVista to the end user.

## METHODOLOGY

HyVista delivers HyMap data as 'radiance at sensor'; therefore the NPD dataset was first treated in order to retrieve the surface reflectance. The effect of absorption by the atmosphere was removed using the ATREM radiative transfer code (Gao, *et al.* 1993). Differences in cross-track illumination caused by the variation in solar angle were removed using a third order polynomial correction provided by the Cross Track Illumination feature in the computer program 'ENvironment for Visualising Images' (ENVI) (http://www.rsi.com). In order to enhance absorption features within the spectra, the dataset was treated with the 'continuum removal' process provided with ENVI (Clark, *et al.* 1987). This process removes the spectral background due to electromagnetic scattering and other effects, thus enhancing the depth of absorption bands.

Georectification, the process of relating each pixel in the dataset to its true location on the ground (eg. correcting for the movement of the aircraft), was carried out using ENVI with flight data files supplied by HyVista. Georectification is carried out as the final step of the analysis since the pixels of the final resultant image are interpolated information rather than 'real data'. After georectification, the multiple flight lines were stitched together as a mosaic using ENVI.

In this project we concentrate on the SWIR region of the spectrum, since this contains distinctive absorption bands directly applicable to hydrothermal minerals. Five library SWIR continuum removed spectra are illustrated in Figure 3. Unique spectral features have been highlighted. These features were used to generate mineral occurrence maps. An automatic search for these features was carried out in each pixel of the hyperspectral dataset, using a similarity algorithm called Spectral Feature Fitting (Clark*, et al.* 1990). This algorithm was chosen because it was found to give best results of all the spectral mapping techniques provided with ENVI (eg. Spectral Angle Mapper, Binary Encoding, etc.). This judgement was based on each algorithm's ability to produce coherent spatial mineral detections and discriminate against 'outliers' (that create a speckled or noisy appearance in the resultant occurrence map).

In addition to spectral mineral mapping, techniques such as examinations of false colour images can help to highlight regions of differing mineralogy. In this project, this has been achieved by examining regions in the 2.3 micron and 2.2 micron regions, and assigning the RGB colour planes to particular HyMap channels in this region. This has the effect of highlighting subtle differences in the absorption bands associated with Mg-OH (2.3 microns) and Al-OH (2.2 microns) vibrations (Hunt 1979).

In order to ground truth the mineral maps, two field seasons have been carried out and representative samples collected. Primary methods of analysis included transmitted and reflected light analysis of petrographic thin sections (PTS) and Electron Microprobe (EMP) analysis of polished thin sections.

## RESULTS

The dataset at the North Pole Dome consists of 14 flight lines that are 2.3km wide on average and from 6 to 22km long. It was collected on October 22, 2002, from 1030-1230 Australian Western Standard (local) Time (AWST). The collection date was chosen to minimize green vegetation and thus was at the end of the tropical dry season. The swathes were flown east to west order, odd swathes flying south, even swathes in a north direction. The average altitude of the aircraft was 2460m above sea level and the average elevation of the terrain was 65m. Representation of the coverage is given in Figure 1. A subset of this area, indicated in Figure 1, has been investigated and is reported on here.

*Mineral Maps*

Figure 4 displays the mineral maps that have been derived for each target mineral. Previous field mapping using airborne photographs has only facilitated the identification of linear features, each thought to be chert veins or faults (Van Kranendonk 2000). This study has shown that in addition to the chert veins, there are two other vein types, one bearing muscovite and the other rich in hornblende. In addition, the presence of a 5-10m thick serpentinised unit above the cherts of the Dresser Formation was previously unrecognized – the unit had been mapped as a carbonated basalt (Van Kranendonk 2000).

The muscovite veins consist of a quartz porphyry where the groundmass has now been completely sericitised. Field checking revealed that most of the veins are of varying width, usually 5-20m across, however their apparent size in the airborne dataset is greatly exaggerated by the position of the veins atop minor topographic protuberances, resulting in the development of scree slopes rich in muscovite splaying out to each side.

As can be seen from Figure 3, some of the minerals have overlapping identifying features, and this makes them hard to discriminate, especially in the presence of noise. Kaolinite and muscovite are hard to discriminate from each other, and if kaolinite is present it is not possible to say definitively whether muscovite is present. We have adopted a position that if there is a kaolinite feature near 2.16, then we cannot say anything about the presence of muscovite; therefore it does not appear in the muscovite map. Serpentine, chlorite and hornblende are also difficult to discriminate due to overlapping features.

*False Colour Images*

The Mg-OH map (Figure 5a) shows chlorite rich zones in blue. A blue band ($\iota$) running north-south at the interface between the Dresser Formation and Apex Basalt is accompanied by several alternating blue and green bands ($\varphi$). These are interpreted to represent separate volcanic flows. The presence of pillows in the Apex Basalt suggests the bands represent submarine lava flows with slightly differing amounts of iron and magnesium, possibly representing geochemically different flows or different amounts of seafloor alteration (Terabayashi*, et al.* 2003). Pervasive argillic alteration at the Dresser Formation appears in white ($\lambda$). An outstanding problem at the moment is why some chlorite-rich regions at the top of the Dresser Formation ($\mu$) escaped this argillic alteration.

The Al-OH map (Figure 5b) shows minor variations in kaolinite and muscovite mineralogy. Intense muscovite alteration is restricted to the quartz porphyry veins and associated scree zones ($\iota$). Some muscovite seems present in the chert at the top of the Dresser Formation. The grey colour over the basalts of the Dresser Formation indicates a weak but pervasive phyllic-argillic alteration that is restricted to the Dresser Formation ($\varphi$). This is interpreted to be associated with the presence of the black chert veins which are ubiquitous throughout the unit. Hydrothermal activity associated with these veins, which do not penetrate the overlying Apex Basalt, is likely to have produced a weak sericitisation throughout the Dresser Formation, which is absent from the Apex Basalt above and the Mt Ada Basalt beneath.

*Ground Truthing and Geochemical Analysis*

Table 4 gives a summary of mineral phases detected from collected field samples. Samples were collected from within spectrally distinctive units, using the hyperspectral dataset as a guide. Hydroxyl bearing minerals phases such as chlorite, hornblende, muscovite, kaolinite and serpentine are all detectable and verified from ground samples.

Ground truthing of the serpentine horizon at the interface of the Dresser Formation and Apex Basalt revealed an ultramafic peridotitic komatiite unit, with olivine replaced by serpentine (Brown, *et al.* 2004).

Minor problems did occur in some instances, particularly in the west (left) of Figure 4, serpentine was mapped in some locations but on ground inspection, only carbonatised basalts were found – carbonate does have an absorption at 2.3 microns which overlaps with serpentine. Work is continuing to address these problems and refine the algorithms used.

Quartz is present in the muscovite bearing veins, however it is not detectable as a mineral phase in the VNIR hyperspectral dataset, therefore it has not been discussed. Feldspars and pyroxenes are generally not discriminable in the SWIR region. The only sulphate mineral in large quantities in this study, barite, does not have a unique spectral feature in the SWIR. Chert is present in veins in the Dresser Formation; however it is relatively featureless in the VNIR-SWIR. Future versions of HyMap with coverage in the thermal infrared (8-12 microns) may be more successful in mapping these minerals.

## DISCUSSION

High magnesian lava flows (komatiites), such as those detected in this study at the base of the Apex Basalt, have been proposed to be common on Mars (Baird and Clark 1984). Komatiite shield volcanoes have been proposed as a possible habitat of early life in the Archaean (Nisbet and Sleep 2001). Remote mapping of serpentine bearing horizons on Earth and Mars, such as demonstrated here using a VNIR-SWIR hyperspectral sensor, may help to shed new light on likely locations for the origin of life on both planets.

Several studies have suggested the possible emergence or preservation of Martian life at sites of hydrothermal activity (Walter and Des Marais 1993; Shock 1997; Varnes, *et al.* 2003). Not all hydrothermal systems marked by alteration should be considered ideal for life. For example, areas of potassic alteration in veins associated with porphyry copper deposits are projected to have formed at 600°C (Sillitoe 1993). This is far in excess of the highest temperatures at which organisms are currently known to grow (Kashefi and Lovley 2003). Epithermal alteration events, over deep seated plutons or on the flanks or distal regions of high temperature hydrothermal sites, such as those that have been imaged at the North Pole Dome, should be far more suitable for nurturing biological activity similar to life as we know it on Earth (Nealson 1997).

In searching for biosignatures in a hydrothermally altered terrain, regions of chief interest include contacts between varying alteration mineralogies, such as veins or lineaments of a specific alteration type that are discordant with the present terrain. Possible targets for hydrothermal activity on the surface of Mars include crater rims (Brakenridge, *et al.* 1985; Cockell and Barlow 2002), volcanic edifices (Farmer 1996), gullies proposed to have been created by hydrothermal activity (Gulick 1998; Harrison and Grimm 2002) and possible shallow intrusions of granitic composition exposed by cratering (Bandfield, *et al.* 2004). A shallow granitic intrusion may provide the best chance of finding epithermal alteration similar to that at the NPD. Scree slopes from elevated areas may enhance the detectability of such deposits, as demonstrated at the NPD.

The detection techniques used in this project can be applied directly to OMEGA and CRISM data. A comparison of the capabilities of OMEGA, CRISM, and HyMap (as configured for this project) is given in Table 5. HyMap does not cover the 2.5-5.0 micron mid-infrared regions of the spectra to be used by OMEGA and CRISM due to strong absorptions in the Earth's atmosphere in this region. Carbonates and sulfates have unique absorption bands in the 2.5-5.0 micron region, and another strong hydroxyl band occurs at 2.7 microns (Brown 2005). Mineral maps of hydrothermal zones on Mars from OMEGA and CRISM data, analogous to those generated by this study, will be used to determine the distribution of hydrothermal activity on Mars, past and present. Such maps will naturally be of great use in planning landing locations for the upcoming NASA Mars Science Laboratory (MSL) and the ESA Aurora project ExoMars rover (Vago*, et al.* 2003).

## CONCLUSION

The results of the hyperspectral mapping of hydrothermal alteration at the North Pole Dome presented herein indicate that hyperspectral imaging is able to provide a synoptic and cohesive assessment of alteration mineralogy over a large area in a relatively short amount of time. Maps such as those in Figures 4-5 can be used to direct ground based studies to identified 'hotspots' such as contacts between differing alteration types, cross cutting alteration veins and concentrations of intense alteration.

The results of this study have revealed new details on hydrothermal alteration within the Dresser Formation of the NPD, and provided a means to separate phyllic, argillic, calcic and propylitic alteration zones. Late stage muscovite and hornblende bearing veins have been identified. In addition, the presence of a thin komatiite layer has been discerned at the base of the Apex Basalt.

This study has demonstrated that:

a. In order to detect muscovite-bearing alteration veins and hornblende-bearing gabbro dykes at the scale of those seen in the NPD, a spatial resolution approaching 5m is sufficient to discriminate propylitic, phyllic, argillic, calcic and serpentinitic alteration zoning within and surrounding the veins.
b. In future VNIR-SWIR hyperspectral surveys of the Martian surface, minerals such as muscovite, chlorite, serpentine, kaolinite and hornblende provide readily discernable mineral markers (should they be exposed) for mapping hydrothermal alteration zones.

## ACKNOWLEDGEMENTS

This research would not have been possible without the generous assistance of the Geological Survey of Western Australia (GSWA). Assistance in the field and reviews contributed by Martin Van Kranendonk and Michael Storrie-Lombardi were very valuable. The CSIRO Division of Exploration and Mining is thanked for the loan of their PIMA II field spectrometer. Jonathon Huntington and Peter Mason at CSIRO are thanked for their assistance with processing and interpretation of spectra. Terry and Peter Cocks at Integrated Spectronics are thanked for their unwaivering support and assistance with field equipment. Thanks to Jeff Byrnes and Guillaume Bonello for thoughtful editing which greatly improved this paper.

| *Age (Ga)* | *Unit* | *General Geology* |
|---|---|---|
| | Euro Basalt | Ultramafic-mafic volcanic flows, with pillows |
| 3.458 | Strelley Pool Chert | Chert-Carbonate-clastic sequence |
| | Panorama Formation | Felsic volcaniclastic suite |
| | Apex Basalt | Ultramafic-mafic volcanics with some chert horizons |
| 3.470 | Duffer Formation | Metamorphosed felsic volcanics |
| | Dresser Formation | Mafic volcanics |
| | Mt Ada Basalt | Mafic volcanis, highly metamorphosed |
| 3.515 | Coonterunah Group | Basalts with some cherty layers |

**TABLE 1 – STRATIGRAPHIC COLUMN OF WARRAWOONA GROUP UNITS PRESENT AT THE NORTH POLE DOME (VAN KRANENDONK, 2000). DATES ARE DERIVED FROM U-PB ISOTOPES FROM ZIRCONS IN THE UNITS AND ARE ACCURATE TO APPROXIMATELY 3 MILLION YEARS FROM THORPE ET AL (1992).**

| *Mineral* | *Standard Formula* | *Typical Alteration class* |
|---|---|---|
| | Phyllosilicates | |
| Chlorite | $(Mg,Al,Fe)_{12}[(Si,Al)_8O_{20}](OH)_{16}$ | Propylitic |
| Muscovite | $K_2Al_4[Si_6Al_2O_{20}](OH)_4$ | Phyllic/Sericitic |
| Kaolinite | $Al_4[Si_4O_{10}](OH)_8$ | Argillic |
| Serpentine | $Mg_3[Si_2O_5](OH)_4$ | Serpentinitic |
| | Amphiboles | |
| Fe-Hornblende | $Ca_2(Mg,Fe)_4Al[Si_7AlO_{22}](OH)_2$ | Propylitic/Calcic |

**TABLE 2 – STANDARD FORMULAE OF HYDROXYL BEARING ALTERATION MINERALS. FORMULAE FROM DEER ET AL. (1992). ALTERATION CLASSES FROM THOMPSON ET AL. (1996).**

| | |
|---|---|
| Spectral Coverage | 0.45 – 2.5 microns (VNIR-SWIR) |
| No. of channels | 126 |
| Operating Height | around 2400m AGL |
| Instantaneous Field of View (IFOV) | 2.5 mrad along track, 2.0 mrad across track |
| Field of View (FOV) | 61.3 degrees (512 pixels) |
| Swathe width | 2.3km |
| Average Ground IFOV (pixel width) | 5m |

**TABLE 3 – OPERATING CHARACTERISTICS OF THE HYMAP VNIR-SWIR SPECTROMETER AS CONFIGURED FOR THIS PROJECT.**

| *Unit* | *Field Description* | *Inferred Mineralogy from HyMap dataset* | *Detected mineralogy from hand sample* | *Analytical Methods used* |
|---|---|---|---|---|
| Apex Basalt | Chloritised basalt (w/ pillows) | Chlorite | Chlorite, epidote, Feldspar (albite), pyroxene (diopside) | PTS, EMP |
| Apex komatiite layer | Ultramafic peridotitic komatiite | Serpentine | Serpentine (antigorite), chlorite, calcite, Magnetite | PTS, EMP |
| Amphibole bearing Veins | Gabbro Dyke | Hornblende/Chlorite | Hornblende (Fe-rich), chlorite, epidote, calcite | PTS, EMP |
| Porphyritic Veins | Quartz Porphyry | Muscovite | Quartz, fine grained muscovite, minor iron oxides | PTS, EMP |
| Dresser Fm Basalt | Chlorite+Sericite Altered basalt | Muscovite/Kaolinite | Quartz, fine grained muscovite, kaolinite, chlorite, calcite | PTS, EMP |

**TABLE 4 – MINERAL PHASES DETECTED WITHIN SPECTRALLY DISTINCTIVE UNITS. UNITS ARE IDENTIFIED ON FIGURE 1A. PTS = PETROGRAPHIC THIN SECTION, EMP = ELECTRON MICROPROBE,**

| | **Instrument and Mode** | | | |
|---|---|---|---|---|
| | **HyMap for Pilbara study** | **OMEGA** | **CRISM Hyperspectral** | **CRISM Multispectral** |
| Approx. Pixel Size | 5m | 300m-4.8km | 18-36m | 100-200m |
| Spectral Coverage | 447-2477nm | 350-5090nm | 400-4050nm | 400-4050nm |
| Channels | 126 | 484 | 570 | 59 |
| Spatial Coverage | ~28x20km | ~90% of Mars | ~11x20km* | ~90% of Mars |

**TABLE 5 – COMPARISON OF HYMAP (AS CONFIGURED FOR THIS PROJECT), AND THE MARTIAN ORBITING HYPERSPECTRAL INSTRUMENTS OMEGA AND CRISM. * THE CRISM HYPERSPECTRAL MODE IS PLANNED TO BE USED AT APPROXIMATELY 3000 LOCATIONS ON THE MARTIAN SURFACE.**

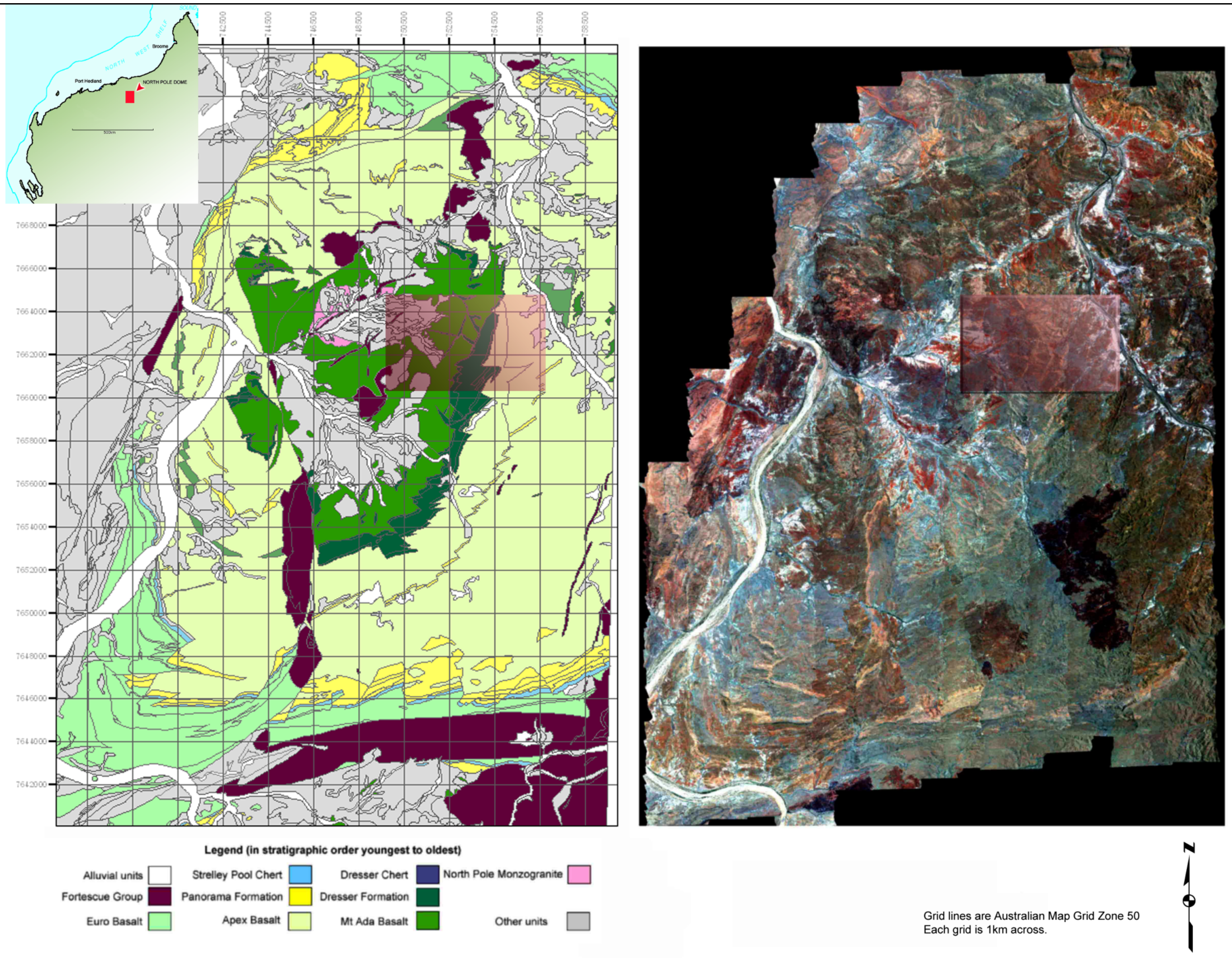

**Figure 1**: (left) – Simplified geological map of the North Pole Dome modified after Van Kranendonk (2000). (right) Visible wavelengths true colour (red: 0.6069 microns, green: 0.5615 microns, blue: 0.4693 microns) image of the hyperspectral dataset. Note 14 swathes have been stitched together and are slightly discontinuous in some regions. Red box indicates the study area reported in this paper.

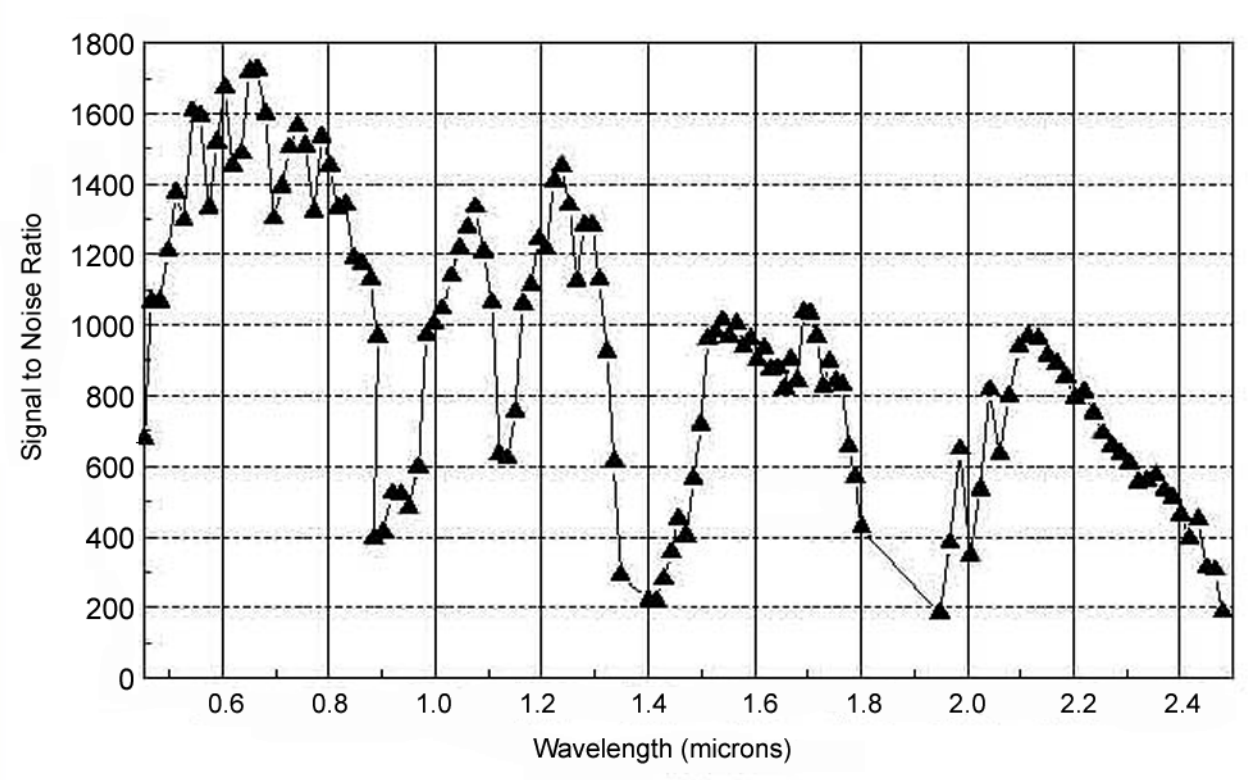

**Figure 2**: Signal to noise ratio of the HyMap instrument. From Cocks *et al.* (1998)

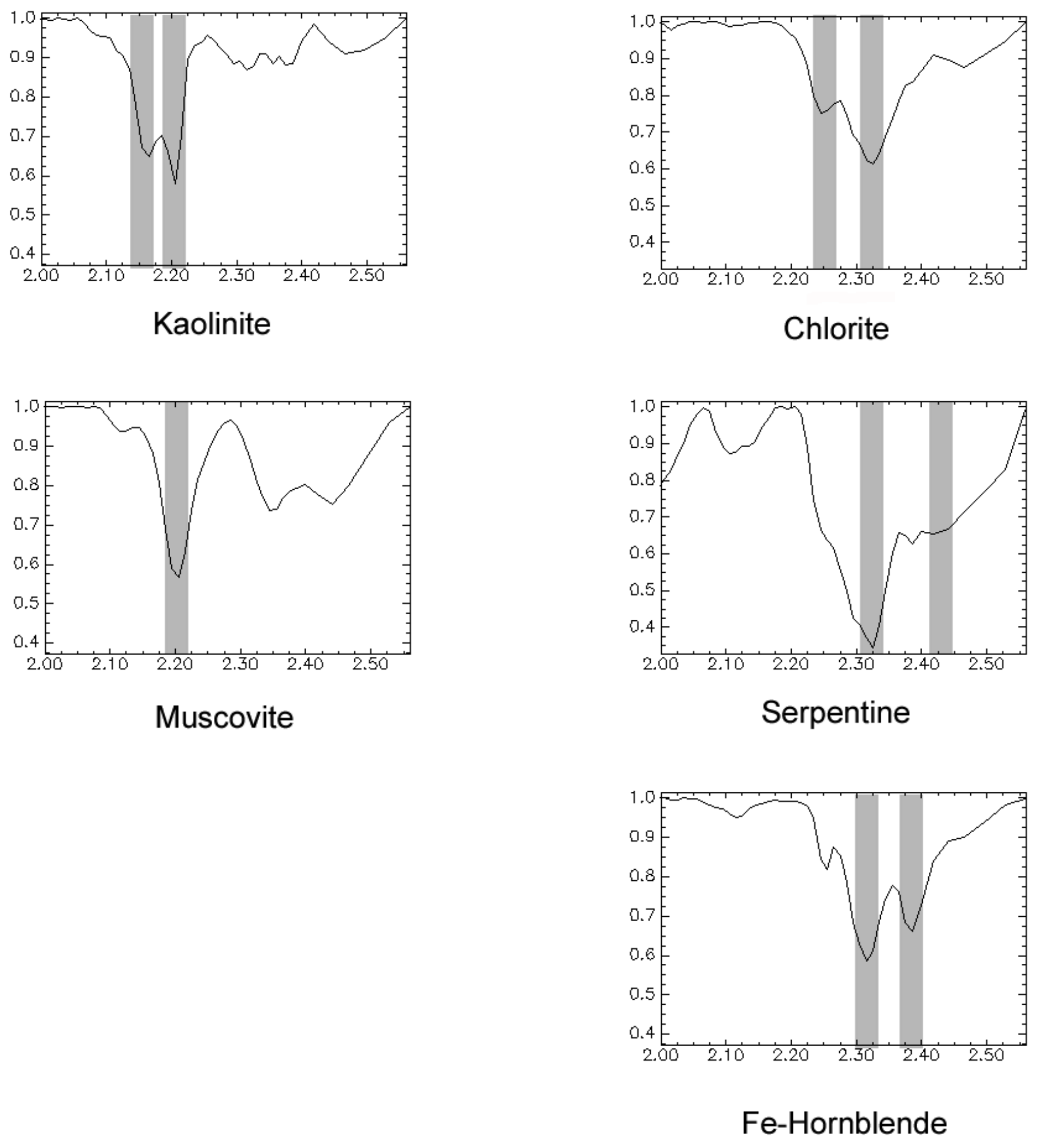

**Figure 3**: Example SWIR spectra of minerals identified in this study, highlighting unique spectral bands used to identify them. Spectra are taken from the USGS Spectral Library described in Clark *et al.* (2003).

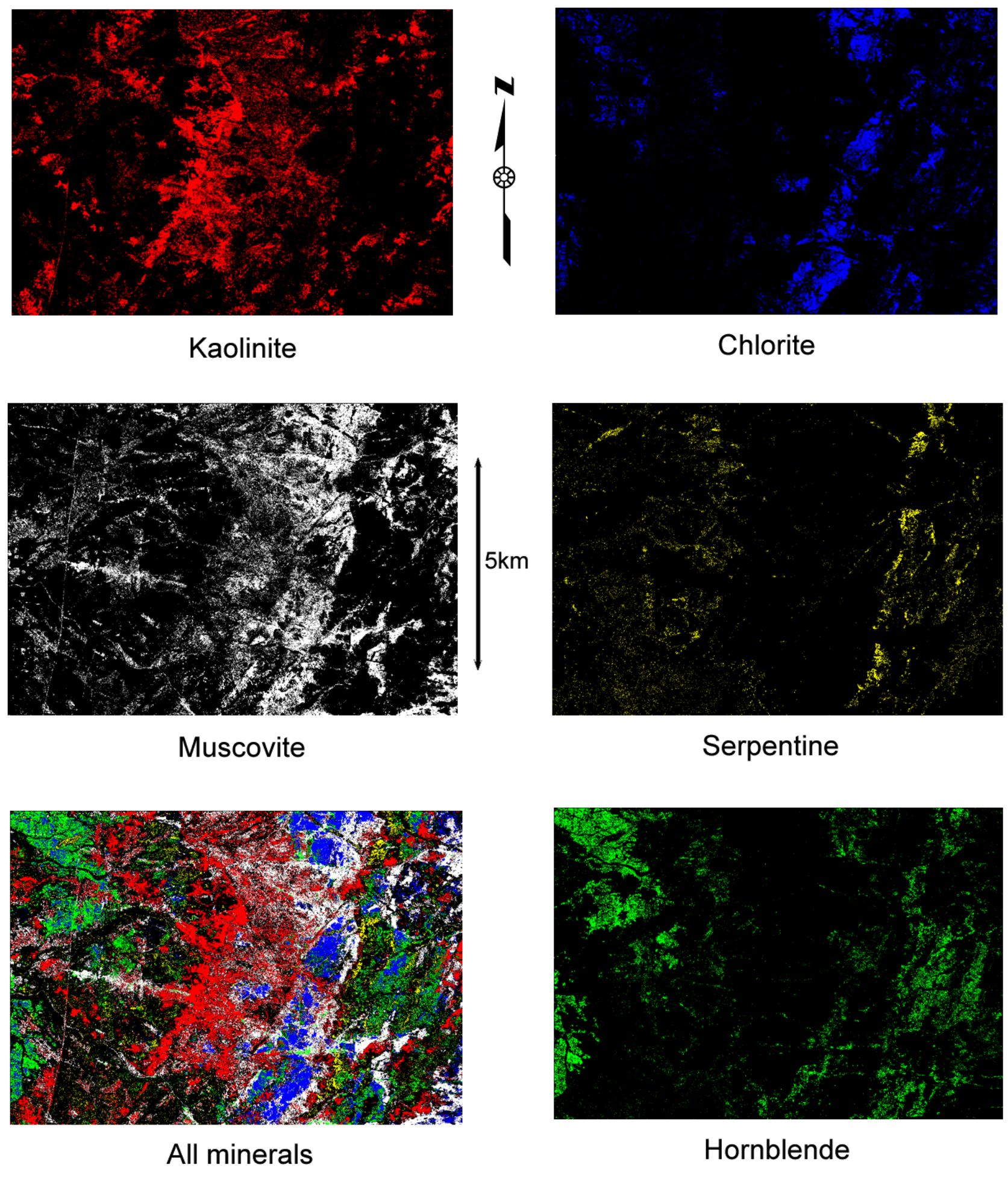

**Figure 4**: Mineral maps of the study region, showing distributions of the indicated minerals. See Figure 1 for coverage. The lower left image is a complete hydrothermal zone map, formed by combining the other five other maps together. Some points are obscured – hornblende (green) was laid down first, then serpentine, then chlorite, then kaolinite, then sericite. This could be viewed as an alteration class map – white is the phyllic zone, red argillic, blue propylitic, green calcic, and yellow serpentinitic.

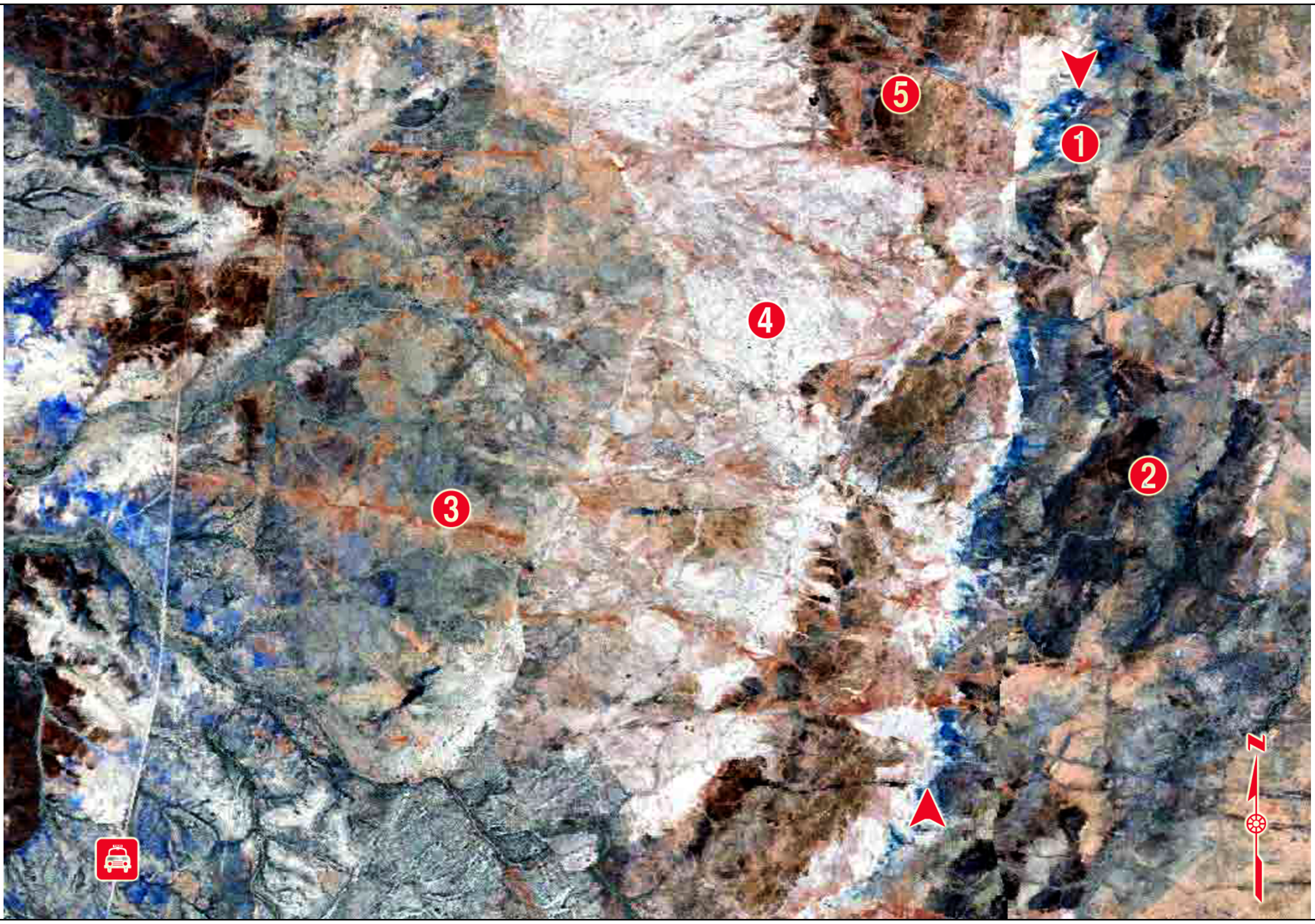

**Figure 5a**: False colour continuum removed image showing variations in Mg-OH (chloritisation) mineralogy in study area. Red is assigned to the HyMap channel at 2.3186 microns, Green to 2.3348 microns and Blue to 2.3515 microns. A thin komatiite layer is in blue, overlying the thin chert layer (in white) at the top of the Dresser Formation, eg. at point ι. The interface between the Dresser Formation and overlying Apex Basalt runs between the two red arrows υ. Overlying chlorites in brown and khaki, eg. at point φ. Muscovite bearing veins are in light brown, eg at point κ. Pervasively altered basalts of the Dresser Formation are in white, as at point λ. There are unaltered Dresser Formation basalts (coloured brown) beneath the white chert layer but above the pervasively altered region, eg. at point μ. The North Pole Road ϕ runs north-south on the left of the image. The image is 5km wide. See Figure 1 for context.

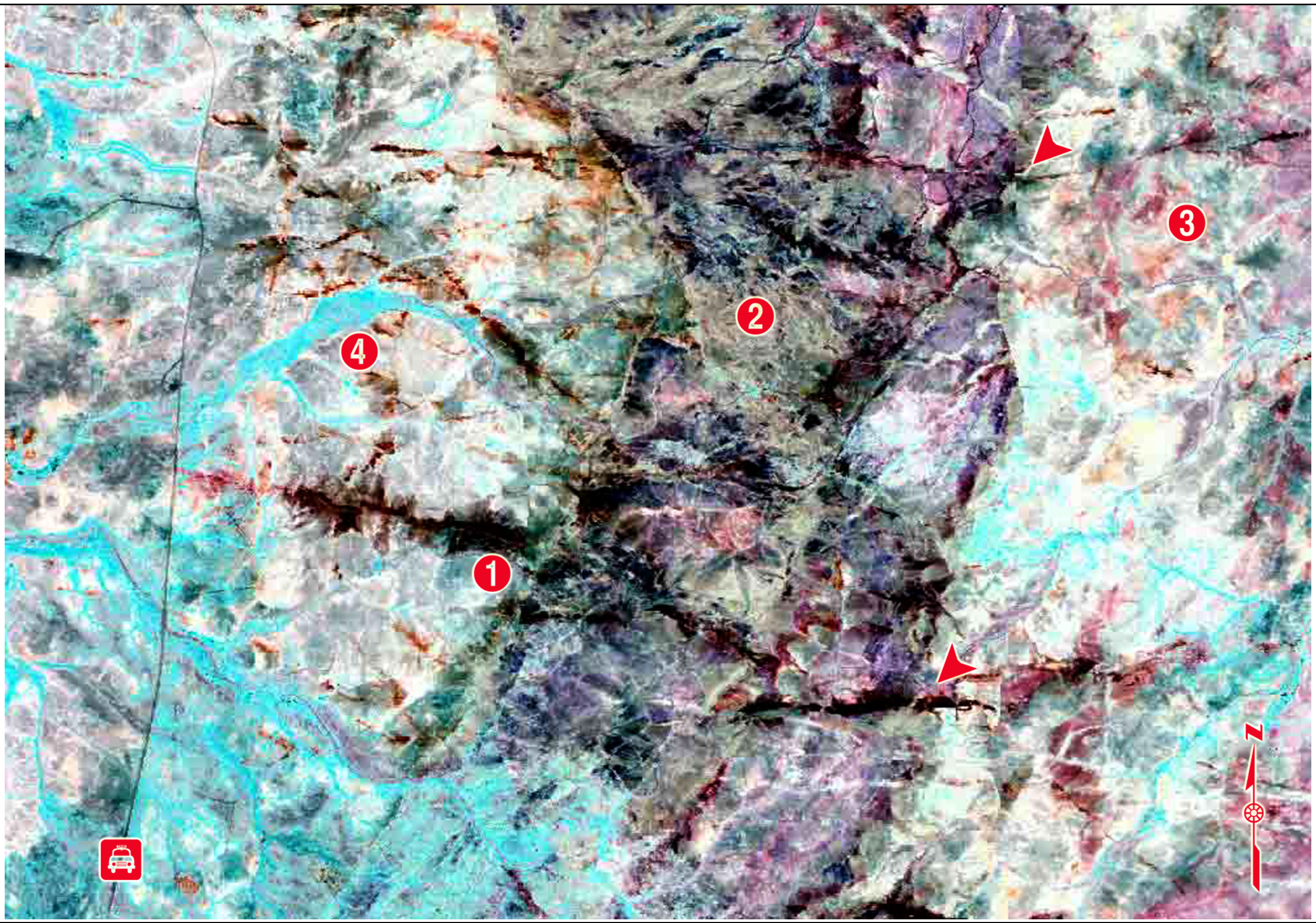

**Figure 5b**: False colour continuum removed image showing variations in Al-OH (muscovite/kaolinite) alteration. Region covered is identical to (a), however the red is assigned to the HyMap channel at 2.1813 microns, Green to 2.1983 microns and Blue to 2.2171 microns. Black indicates intense sericitisation along veins (eg. at point ι), lighter colours indicate little to no white mica present. Grey areas highlight pervasive alteration of the Dresser Formation basalts (eg. at point φ. Lighter regions with no sericite alteration constitute the Apex Basalt on the right of screen (eg. at point κ). Note that at least two muscovite bearing veins penetrate the interface between the Dresser and Apex Formations (at arrows υ). Alluvial fans are highlighted in aqua, eg. at point λ. The North Pole Road ϕ runs north-south on the left of the image. The image is 5km wide. See Figure 1 for context.